\documentclass{elsart}
\journal{New Astronomy}

\usepackage{natbib}

\usepackage{graphicx}
\usepackage{epsfig}

\usepackage{amssymb}

\input psfig.sty
\def\H7{\mbox {$h_{0.7}$}}

\def\IZw18{I~Zw~18}
\def\m82{M82}

\def\h{\mbox {\rm H}}

\def\angs{\mbox {~\AA}}
\def\lya{\mbox {Ly$\alpha$~}}

\def\h0{\mbox {~H$_0$}}
\def\q0{\mbox {~q$_0$}}
\def\o3hb{[OIII]$\lambda5007$~/~H$\beta$~}
\def\O1ha{[OI]$\lambda6300$~/~H$\alpha$~}

\def\s2ha{[SII]$\lambda\lambda6717,31$~/~H$\alpha$~}
\def\2z2{HeII~$\lambda4686$~}
\def\z7{[NII]~$\lambda6583$ }
\def\N2{[NII]~$\lambda6583$~/~H$\alpha$~}
\def\16z2{[SII]~$\lambda\lambda6717, 6731$ }

\def\asec{\ifmmode {'' }\else $''~$\fi}  
\def\amin{\ifmmode {' }\else $'~$\fi}    
\def\arcsper{\ifmmode \rlap.{'' }\else $\rlap{.}'' $\fi} 
\def\arcmper{\ifmmode \rlap.{' }\else $\rlap{.}' $\fi} 
\def\sles{\lower2pt\hbox{$\buildrel {\scriptstyle <}
   \over {\scriptstyle\sim}$}} 
\def\sgreat{\lower2pt\hbox{$\buildrel {\scriptstyle >}
    \over {\scriptstyle\sim}$}} 
\def\flux{~ergs~s$^{-1}$~cm$^{-2}$}

\def\cm3{~cm$^{-3}$}
\def\mpc3{~Mpc$^{3}$}
\def\mpc-3{~Mpc$^{-3}$}

\def\et{{\rm et\thinspace al.}\ }   

\def\apj{ApJ}

\def\aj{AJ}

\begin{document}

\begin{frontmatter}



 \title{Preliminary Results from a Spectroscopic \lya Survey
at Redshift 5.7 with IMACS}


\author{Crystal L. Martin$^{1,4,5}$, Marcin Sawicki$^{1,3}$, Alan Dressler$^2$, and
Patrick J. McCarthy$^2$}
\address{$^1$UCSB, Department of Physics, Santa Barbara, California 93106\\
$^2$OCIW, 813 Santa Barbara Street, Pasadena, CA 91101-1292 \\
$^3$ HIA, 5071 West Saanich Rd., Victoria BC, V9A 7M7 Canada \\
$^4$ Packard Fellow\\
$^5$ Alfred P. Sloan Fellow\\
E-mail: cmartin@physics.ucsb.edu}

\begin{abstract}
  We describe preliminary results from an ultra-sensitive, spectroscopic
emission-line survey and illustrate the challenges inherent in identifying
high-redshift \lya emitters.  Our multi-slit windows technique complements
other types of emission-line surveys. Narrowband imaging surveys cover large
areas of sky but only detect much brighter objects. Longslit spectra taken
along cluster caustics yield intrinsically fainter lensed \lya\ emitters but
probe small volumes of space. We have observed the COSMOS deep field and
a field at 15h +00. To a line flux of a  $few \sim
\times10^{-18}$\flux, we found 150 emission-line sources (with no
detectable continuum) among 4 masks. These candidates 
are being re-observed with broad spectral coverage to determine the line
identity. To date, the interloper to \lya\ ratio is about 8:1.  The
sky positions of the \lya\ candidates generally do not coincide with those
of foreground objects in ultra-deep $r$ band or $i^{'}$ images -- consistent 
with the presence of a strong Lyman break.  
\end{abstract}


\end{frontmatter}

\section{Introduction to the Multislit Windows Technique}

Finding distant galaxies is a contrast problem. The night sky
is much brighter than the galaxies.  Since atmospheric OH emission 
lines dominate this background, ground-based observations are
most sensitive at wavelengths falling between the molecular
line complexes.  Imaging through one of the larger {\it windows} at 
8200\angs\ led to to the first samples of \lya-selected galaxies
at redshift 5.7 (\cite{Hu-LF}, \cite{LALA03}, and \cite{Taniguchi03}).

In principle, the $\sim 150$\angs\ of sky in this OH-free window
can be dispersed with a spectrograph to improve the line-to-sky
contrast by reducing the sky brightness by another factor of 50 
(i.e. 150 \angs / 3 \angs\ FWHM).  Although a number of high-redshift \lya\ 
emitters have been serendipitously discovered during deep spectrosopic 
observations of other targets, the volume probed in a typical longslit 
observation is insignificant.  Custom-designed, multi-slit masks can be 
employed, however, when spectroscopic observations are made with a
band-limiting, OH-suppression filter.  This multi-slit windows technique
has been used previously to search for z=5.7 and z=6.5 \lya\ emitters on 
the world's largest telescopes (\cite{CL99}, \cite{Stockton99}, \cite{MS04}, 
and \cite{TLCB04}). These surveys failed to produce bona-fide \lya emitters
because they did not probe large enough volumes of space. 

Order of magnitude improvements in search volume (at z=5.7) were
realized with the introduction of the IMACS short (f/2) camera on 
the Magellan telescope. We carried out blind \lya\ searches at z=5.7 in 
2004 April, 2005 March, and 2005 May with paired follow-up runs.
The total detector area in spectroscopic mode is 737 square arcminutes; 
approximately 10\% of the field is subtended by the 100 slits, each of width 
1.5 arcseconds, on our masks. Table~1 summarizes the observations to date.
Fields at 10h and 15h have each been observed twice with interlaced masks.
The best data were obtained on the 2005 March run when 36,000~s were obtained 
on the 10h~Field in 0.8 arcsecond, or better, seeing. Sensitivities are
being computed individually for each run, but typical values are a 
$few \times 10^{-18}$\flux.

\begin{table}
\caption{Summary of Blind \lya Searches with IMACS}
\begin{tabular}{llll}
\hline
Date & Field & Integration Time & Conditions \\
  & & (s) & (arcseconds)\\
\hline
\hline
2004 April & 10h Field & 22,800 & clear; 0.8 to 2.5 seeing \\
''         & 15h Field & 39,200 & clear; 0.8 to 2.5 seeing \\
2005 March & 10h Field & 36,000 & clear; 0.6 to 0.8 seeing\\
2005 May & 15h Field & 24,300 & partly cloudy; 0.8 seeing \\
\hline
\end{tabular}
\end{table}

\section{Results from Blind Emission-Line Search}

The reduced, sky-subtracted data were searched for emission lines.
If continuum emission was detected blueward of the discovery line, 
then the line was rejected from the list of \lya\ candidates. 
Intergalactic absorption toward a redshift 6 galaxy effectively
moves the Lyman break from 912\angs\ up to the \lya\ line \cite{Madau95}.
Many \lya\ candidates were found in each field, and a subset of these
were selected for follow-up spectroscopy without the narrowband blocking
filter.  Table~2 shows the yield of total emission-line objects, \lya
candidates, and confirmed \lya\ emitters to date.

\begin{table}
\caption{Emission-Line Objects Discovered in Blind Search} 
\begin{tabular}{lllll}
\hline
Date & Field & N (Emission Lines) & N (\lya Candidates) & N (\lya Confirmed) \\
\hline
\hline
2004 April & 10h Field & 87 & 52 & 3 \\
``         & 15h Field & 86  & 48 & 5 \\
2005 March & 10h Field & 67 & 31 & 6\\
2005 May & 15h Field & 25 & 22 & TBD\\
\hline
\end{tabular}
\end{table}

\section{Confirmation via Spectroscopic Follow-Up}

\subsection{10h Field}

Follow-up spectra of the 10h Field were obtained 2005 May.
Slits were placed on 43 targets from the 2004 April search
and 48 targets from the 2005 March search. We re-detected
all but 3 of the 2005 March objects.  The positions for
the 2004 April candidates are less accurate than those for
the 2005 runs. The addition of the Magellan corrector lens 
in the middle of 2004 changed the focal length, and the scale 
changes in the mask plane have not been accurately modeled. 
Among the 2004 April targets, only 28 of 43 were recovered. 

We identified about 8 interlopers per \lya (and [OII]) emitter.  The
line was positively identified for about one-third of the interlopers.
A \lya\ identification was ruled out for the other two-thirds due to
presence of continuum emission blueward of the discovery line and/or 
the presence emission-lines blueward of the discovery line.

\begin{figure}[!t]
\centerline{\psfig{file=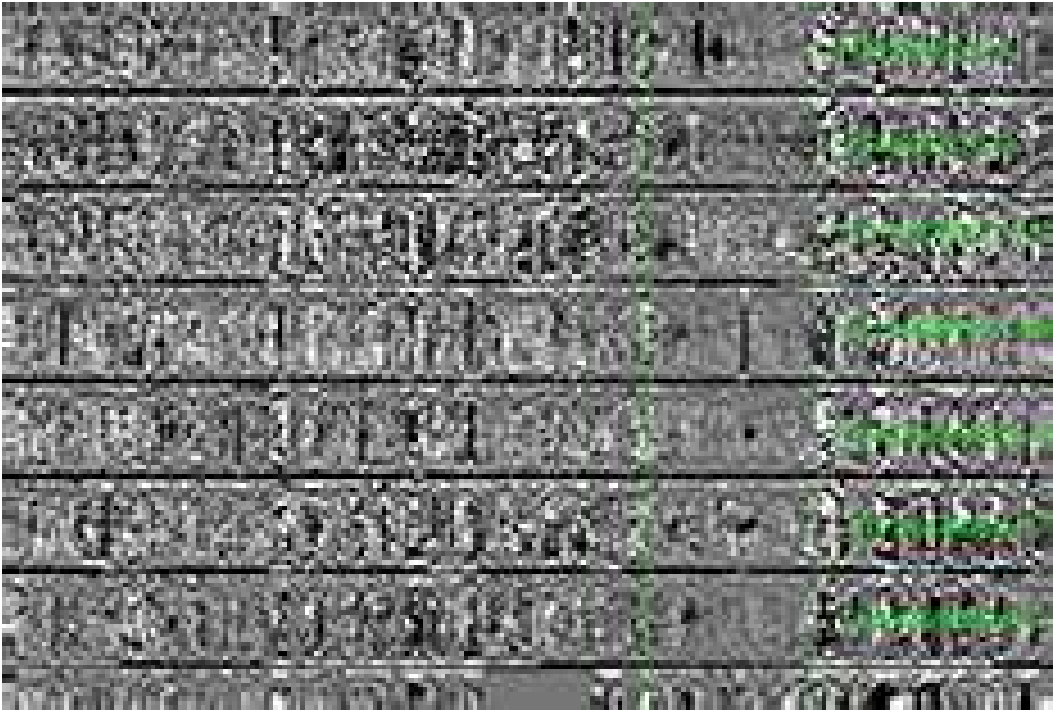,width=5.5in}}
\caption{Confirmation spectra of several \lya emitters from the 10h Field.
Spectra are aligned in wavelength, and the {\it discovery lines}  can
be seen in the low background window just the right of center at 8200\angs.
The candidates are, from top to bottom, 
MSDM30.5-8.1,
MSDM17.5+8.1,
MSDM25.5+5.1,
MSDM75.5-6.1,
MSDM100.5-2.1,
MSDM26.5+7.1,
MSDM80.0+3.1,
MSDM64.0-5.1,
and MSDM52.0+8.1.  
(One object, MSDM52.0+8.1, is not shown due to resolution limits on the 
display.)
}
\label{fig:10h_lyaspec}
\end{figure}

Inspection of the spectra left 6 and 3 likely \lya\ emitters
from the 2005 March and 2004 April candidates, respectively. 
Figure~\ref{fig:10h_lyaspec} 
shows the spectra. Our follow-up spectra had to be taken at lower
resolution than originally planned, and the [OII]3726,29 doublet is 
not resolved. Our sample of \lya\ emitters may include several [OII] 
interlopers. The follow-up spectra  are shown in Figure~\ref{fig:10h_lyaspec}.

\subsection{15h Field}

Follow-up spectra were obtained in 2004 July for the 15h Field. Slits
were assigned for  37 of 48 objects from the \lya-candidate list. (The
others had overlapping spectra or discovery lines in the inter-chip gaps.)
We detected 16 of these emission-line objects. The positions of candidates 
in the mask plane could not be modeled as accurately as desired because 
the corrector was installed between the discovery run and the follow-up run. 
The quality of the discovery line in the un-detected candidates was as high 
as that of the re-detected candidates. Apparently the model
of the  scale changes in the mask plane was not perfect.

\begin{figure}[!t]
\centerline{\psfig{file=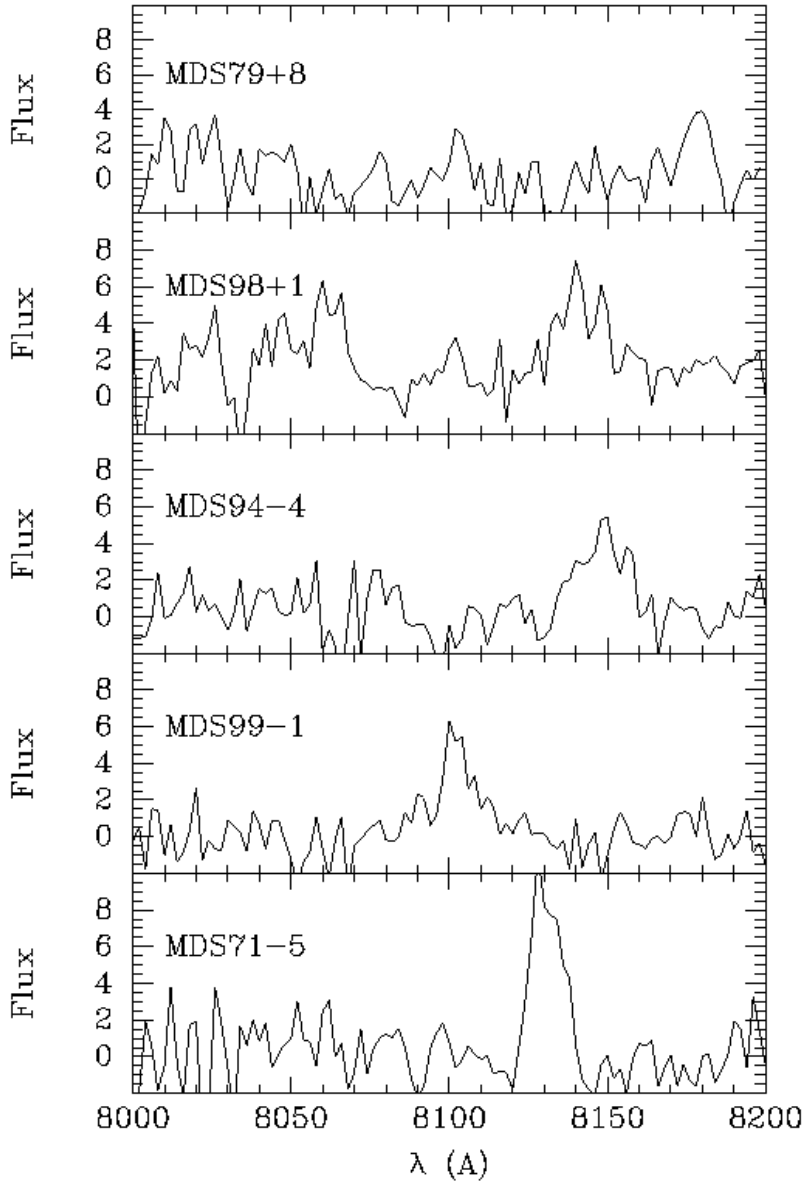,width=4.5in}}
\caption{Follow-up spectra of 15h Field \lya\ candidates from
2004. Confirmed \lya\ (or [OII]) emitters include MSDM79+8,
MSDM98+1, MSDM94-4, MSDM99-1, and MSDM71-5.}
\label{fig:15h_lyaspec}
\end{figure}

We identified 8 of these 16 objects as foreground galaxies based
on their spectra. One, MSDM94-8, was identifed as foreground based
on a positional coincidence with a foreground galaxy.  Five 
objects were confirmed as \lya\ (or [OII]) emitters, and these
are shown in Figure~\ref{fig:15h_lyaspec}. The two remaining objects
have not been classified yet.

\section{Confirmation via Continuum Break}

Any high-redshift \lya\ emitter must present a strong
continuum break shortward of the \lya\ line \cite{Madau95}.
Faint, foreground galaxies with an interloping emission-line
in the 8200\angs\ OH-window will normally have fairly blue colors.
Broadband imaging shortward of the 8200\angs\ window therefore
provides another check on the identity of the $\sim 8200\angs\ $ line.
True \lya\ emitters will not be detected in this  {\it veto image}.

We obtained an ultra-deep Subaru $r$ band image of the 10h field  from 
the COSMOS team and an $i^{'}$ band image from the CFHT Legacy Survey.
The sky positions of our candidates were computed from a transformation
calculated by Ken Clardy.  For the 2005 data on the 10h field,
our experiments show that the positional uncertainty along the slit is 
less than 0.5 arcseconds. Perpendicular to the slit, the error is
assumed to be half the slit width, or 0.75 arcseconds. The uncertainty 
for the 2004 positions in both fields may be somewhat larger.

\begin{figure}[!t]
\centerline{\psfig{file=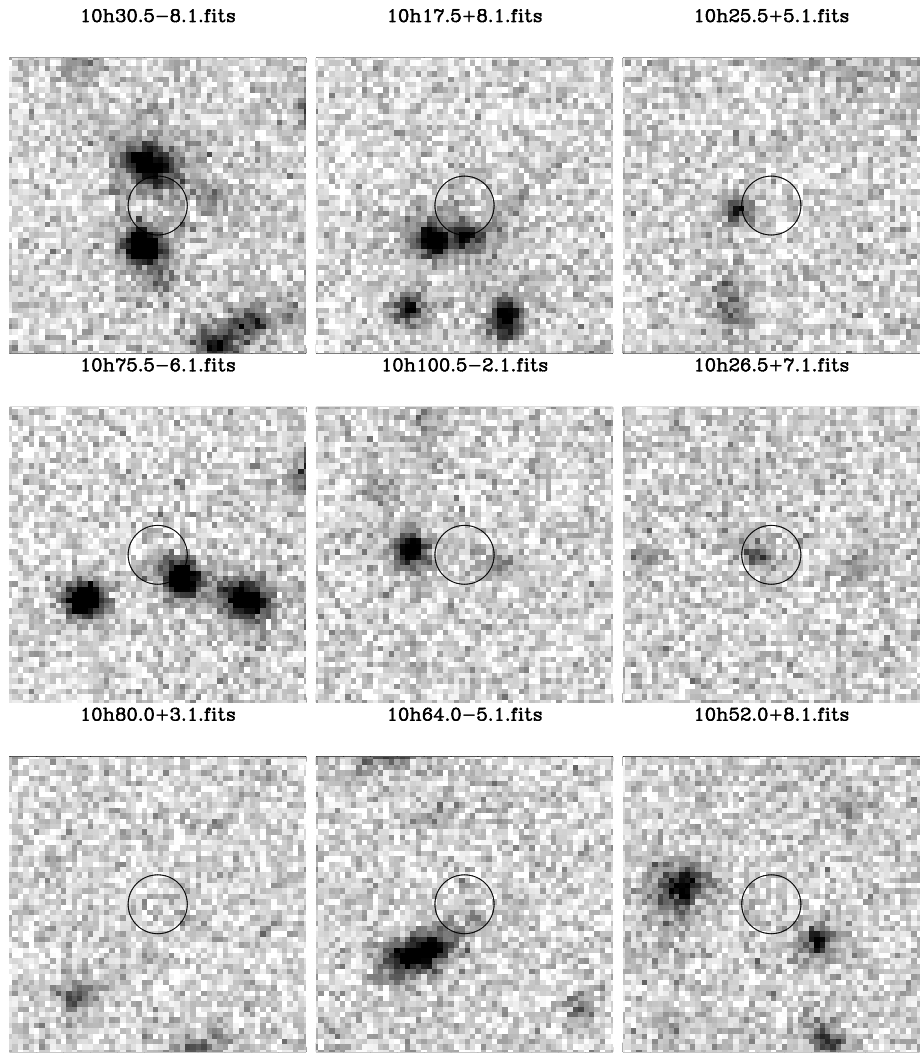,width=5.5in}}
\caption{Position of \lya candidate on ultra-deep $i^{'}$ band image
of the 10H field.
North is up and east is to the left. The circles are 1\asec in radius,
which is about  twice the positional uncertainty. Each image is 10\asec\
on a side; north is up and east is to the left.}
\label{fig:10hcand}
\end{figure}

Figure~\ref{fig:10hcand} shows the positions of the \lya\ candidates from 
the 10h field on the CFHT image. Only object MSDM26.5+7.1 is clearly 
associated with an object in the $i^{'}$ image.  {\it We find more 
\lya\ candidates within 2 arcseconds of foreground galaxies than 
expected.} Some fraction of these are probably  [OII] emission
lines from HII regions in the outer parts of the foreground galaxy. Spectra
of the foreground galaxies or higher-resolution spectra of the candidates
should easily determine which objects are interlopers.

A deep $B$ band
image of the 15h field was obtained previously with the NOAO mosaic camera.
Positional errors should be similar to those for the 10h field.
In the 15h field, only MSDM94-8 was rejected due to an exact positional
coincidence with a foreground object.

\section{Role of \lya - Selected Galaxies in Reionzation}

Measurement of the luminosity distribution of \lya\ emitters is critical
to understanding the role of emission-line selected galaxies in maintaining
the ionization of the intergalactic medium at redshift 6 \cite{MS04}. A full
analysis of the data presented here should  provide the best constraints
on the faint-end slope to date.   Existing observations that probe
slightly fainter emitters probe such small volumes that
the uncertainties in the number density are large (\cite{ESKK01}, 
\cite{Santos04}). 
The strong clustering properties of the \lya\ emitters have been emphasized
by \cite{Hu-LF}  in their estimate of the luminosity distribution. 
In addition, the evolution, or lack of evolution,
in the \lya\ luminosity function at this epoch is an important constraint
on the epoch and progression of reionization \cite{MR04}. 

At present, we are working to finish the data processing for the 15H Field
follow-up.  We can also investigate the proximity of the \lya\ candidates
to foreground galaxies three ways:  spectra of the foreground objects,
higher resolution spectroscopy, and Monte Carlo simulations.  The effective
survey volume is a function of object flux, and this function is being
determined with simulations.  With these results in hand, we will compute
the faint-end slope of the \lya\ luminosity function at $z=5.7$.

The authors would like to thank Yoshi Taniguchi, Peter Capak, and Patrick
Shopbell for making the subaru r band image of the COSMOS field available
and the CFHTLS team for the  $i^{'}$image of the CFHTLS 10h field. Ken
Clardy contributed enormously to modeling the transformation from the
mask plane to the sky and back, and his thoughtful help is greatly appreciated.

\end{document}